# Humidity Effect on Diffusion and Structure of a Moisture-Swing CO$_2$ Sorbent

Xiaoyang Shi,[1] Hang Xiao,[1] Xi Chen,[1] Tao Wang[2], Klaus. S. Lackner[3]

[1] Department of Earth and Environmental Engineering, Columbia University, New York, NY 10027, USA

[2] State Key Laboratory of Clean Energy Utilization, Zhejiang University, Hangzhou, Zhejiang, 310027, China

[3] School of Sustainable Engineering & Built Environment, Arizona State University, Tempe, AZ 85287-9309, USA

## Abstract:

Ion hydrations are ubiquitous in natural and fundamental processes. A quantitative analysis of a novel CO$_2$ sorbent driven by ion hydrations was presented by molecular dynamics (MD). We explored the humidity effect on the diffusion and structure of ion hydrations in CO$_2$ sorbent, as well as the working mechanism of the moisture-swing CO$_2$ sorbent. The discovery of the diffusion coefficients of ions at different humidity levels can be applied to design a new CO$_2$ sorbent with superior performance. The molecular structure analysis shows a higher precipitation rate of HCO$_3^-$ ions than CO$_3^{2-}$ ions in a dry surrounding, which leads to a more alkaline surrounding on CO$_2$ sorbent that may promote the absorption of CO$_2$.

# 1 Introduction

Hydration of neutral and ionic species at interfaces plays an important role in a wide range of natural and fundamental processes, including in chemical systems[1,2] as well as biological[3] and environmental systems[4]. The hydration of ions affect the physical structure and the chemical energy transfer through the formation of highly structured water[5-10]. Of particular interest in this study is a novel technology for direct air capture of $CO_2$[11], driven by the free energy difference between the hydrated and dehydrated states of an anionic exchange resin (IER)[1]. This material doped by carbonate solution can absorb $CO_2$ from ambient air when surrounding is dry, while release $CO_2$ when surrounding is wet, as depicted in **Equation 1-4** and **Figure 1**. **Equation 5** depicts the reactions of anions with the change of the number of water molecules (n) on IER. When the surrounding is dry (n is small), the reaction moves to right side which produces large amount of $OH^-$ ions to absorb $CO_2$, while when the surrounding is wet (n is large), the reaction moves to left side preferring $CO_3^{2-}$ ions in the system[1].

$$H_2O \Leftrightarrow H^+ + OH^- \qquad (1)$$

$$CO_3^{2-} + H^+ \Leftrightarrow HCO_3^- \qquad (2)$$

$$OH^- + CO_2 \Leftrightarrow HCO_3^- \qquad (3)$$

$$HCO_3^- + HCO_3^- \Leftrightarrow CO_3^{2-} + CO_2 + H_2O \qquad (4)$$

$$CO_3^{2-} \cdot nH_2O \Leftrightarrow HCO_3^- \cdot m_1 H_2O + OH^- \cdot m_2 H_2O + (n - 1 - m_1 - m_2)H_2O \qquad (5)$$

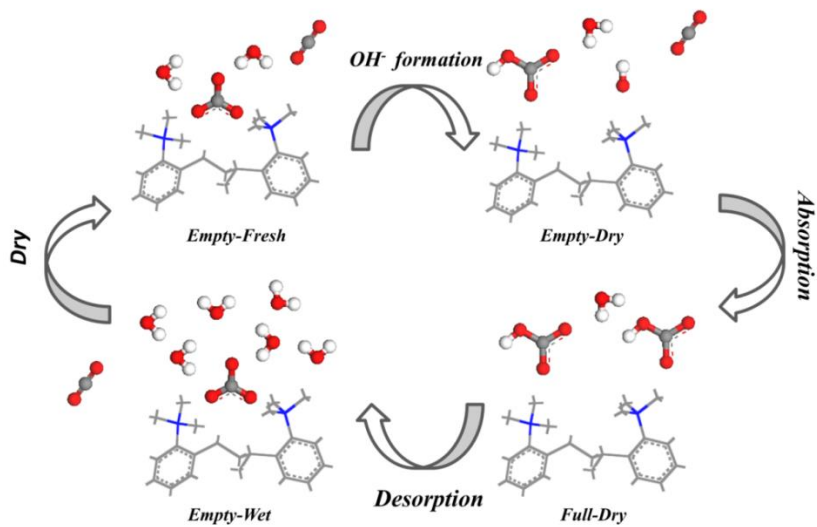

Figure 1: Reaction pathway of CO2 absorption/desorption on IER. ***Empty-Fresh*** state is the sorbent in dry condition with a few water molecules in the surrounding. ***Empty-Dry*** state is $H_2O$ splits into $H^+$ ion and $OH^-$ ion which is ready to absorb $CO_2$, and $H^+$ ion is combined with $CO_3^{2-}$ forming $HCO_3^-$ ion. ***Full-Dry*** state is the full-loaded sorbent in the dry condition. The three states (**Equation 1-3**) present the absorption process. ***Empty-Wet*** state is regenerate absorbent releases $CO_2$ in the wet condition (desorption **Equation 4**).

The quaternary ammonium cations are attached on the backbone of IER, while $H_2O$ molecules and three types of anions $CO_3^{2-}$, $HCO_3^-$, $OH^-$ are moveable by different diffusion rates. The diffusions of movable ion hydrations and water molecules on IER are essential to determine the absorption efficiency of the sorbent. By studying the diffusivities and structures of functional substances under different moisture concentrations, helpful insights may be deduced to understand the underlying working mechanisms and to design a more efficient sorbent for $CO_2$ capture from ambient air.

MD simulations are especially appropriate for studying complex polymer systems[12-16] and water structures[10,17,18], since it can be applied to expose nano-structure features without a priori structural model. Researchers have calculated the diffusion of molecules in polymer system[19-23] and investigated the moisture effect on epoxy resins by MD simulation. Lin[24] investigated the diffusion coefficient and the activation energy of epoxy resin under moisture environment and showed that the results from MD simulations and experiments have an acceptable difference. Wu[25] studied the influences of absorbed water on structures and properties of crosslinked resin including the diffusion coefficient of water, radial distribution function, geometry configuration and mobility of polymer network chains. Chang[26] performed MD to study the hygroscopic properties of resin materials regarding diffusivity and swelling strains respect to temperature and moisture concentration. Lee[27] simulated the distribution and diffusion of water in epoxy molding compound, considering the effect of water content.

Although diffusion of moisture in polymer has been studied by experiments and computer simulations, the transport properties of anion exchange resin (IER) for moisture-swing $CO_2$ capture sorbent in air surrounding have not been involved in research, since IER was previously used for water treatment [28-30]. Since the successful demonstration of applying IER for $CO_2$ capture from ambient air [11], thermodynamic[31] and kinetic[32] investigations were carried out to explain the underlying moisture-swing mechanism[1,33,34]. Nevertheless, the diffusive and transport characteristics remains unclear at molecular level. In this study, for the first time, MD simulation was carried out to investigate the diffusivity and structures of ions and water molecules in a $CO_2$ capture sorbent under various humidity conditions.

# 2 MD Simulation

## 2.1 Models of Ion Exchange Resin

The IER in the simulation is composed of a polystyrene backbone with quaternary amine ligands attached to the polymer. These quaternary amine groups carry a permanent positive charge. They can be depicted as $NR_4^+$, in which R is an organic carbon chain attached to the polymer matrix. The positive ions fixed to the polymer backbone do not release proton.

A model of oligomer containing eight side chains with eight quaternary ammonium ions was established for MD simulation. A oligomer includes two quaternary ammonium ions is shown in **Figure 2**. Four oligomers, each containing eight quaternary ammonium ions, were packed in an amorphous cell. The periodic boundary conditions was applied to eliminate surface effects. In this study, two IER systems containing different classes of anions were established in charge balance. System 1 has four oligomers[35] attaching sixteen carbonate ions, and the other one system 2 has four oligomers attaching sixteen bicarbonate ions and sixteen hydroxide ions. System 1 and system 2 represent the reactant and product of **Equation 5,** respectively, shown in **Figure 3**.

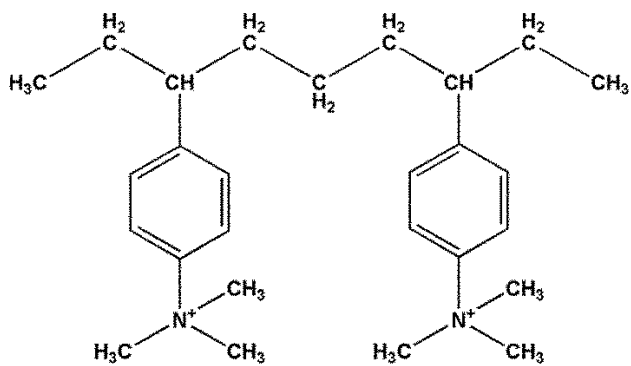

**Figure 2:** Chemical structure of IER containing two side chains

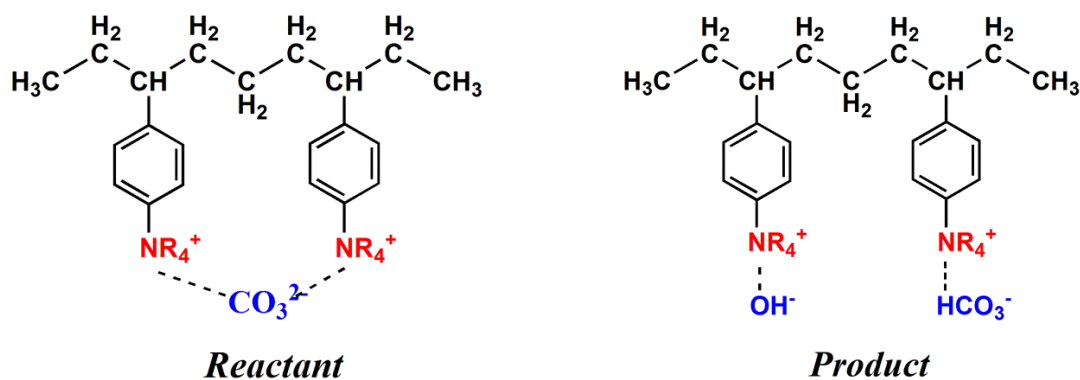

**Figure 3** Chemical structures of reactant system 1 and product system 2

System 1 (S1) and system 2 (S2) are solvated with different numbers of water molecules. In the carbonate ion system (S1) simulations, the $CO_3^{2-}$ : $H_2O$ ratio is selected to be 1:5, 1:10 and 1:15 (total water molecule numbers are 80, 160 and 240 in the computational cell) respectively, and for the bicarbonate and hydroxide ion system (S2), $HCO_3^-$ : $H_2O$ ratio or $OH^-$ :$H_2O$ ratio is tested at 1:4, 1:9 and 1:14 (total water molecule numbers are 64, 144, and 244 in the computational cell) respectively, from low to high humidity conditions. These cases have one-to-one correspondence, since one water molecule reacts with one carbonate ion to form a bicarbonate and a hydroxide ion. The geometry configurations of S1 containing 80 water molecules and S2 containing 64 water molecules are shown in **Figure 4,** which have been optimized for MD calculation.

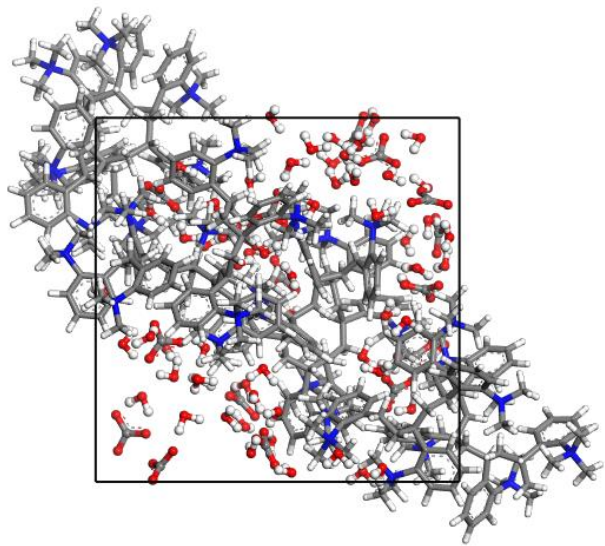

(a)

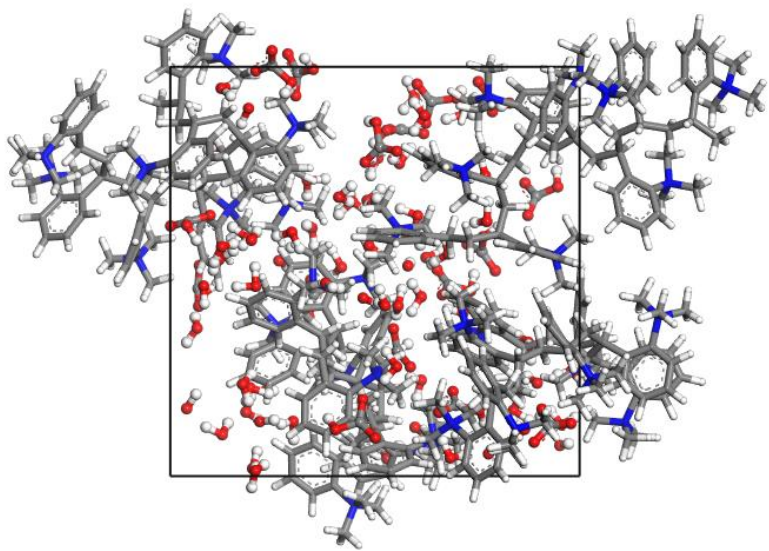

(b)

**Figure 4:** Geometry configurations of IER with ion species and water molecules. **(a)** S1 contains 4 oligomers, 32 quaternary ammonium ions, 16 carbonate ions, and 80 water molecules. **(b)** S2 contains 4 oligomers, 32 quaternary ammonium ions, 16 bicarbonate ions, 16 hydroxide ions and 64 water molecules.

## 2.2 Simulation Procedure

All molecular dynamics simulations were carried out in Materials Studio,[36] COMPASS Force Field was used for all geometry optimizations and MD simulations. COMPASS uses an *ab initio* force field optimized for condensed-phase applications. This force field was assigned to all atoms in the carbonate ion, bicarbonate ion, hydroxide ion, and water molecule.

The initial six structures of S1 and S2 with three different numbers of water molecules each were build in amorphous cells. Minimizations were carried out by the Quasi-Newton procedure, where the electrostatic and van der Waals energies were calculated by the Ewald summation method (the Ewald accuracy was 0.001kcal/mol, and the repulsive cutoff for van der Waals interaction was 6 Angstrom). In order to achieve a relaxed structure, the systems were further equilibrated by NVE ensemble simulation with 100 ps, and then an NPT ensemble was performed to obtain the relevant density values with different water numbers at standard state condition. The system achieved equilibrium by running 200 ps in NVT ensemble. To estimate the diffusivity and structure of molecules, NVT ensembles for 0.5 ns were run with different densities at 298 K. A time step of 1.0 fs was used in all simulations. NPT ensemble used Nose thermostat and Berendsen barostat, and NVT ensemble used Nose thermostat.

The diffusion coefficients for all ions and molecules were calculated from the Einstein relation[37] as **Equation 6**

$$D = \frac{1}{6N} \lim_{t \to \infty} \frac{d}{dt} \sum_{i=1}^{N} \langle |r_i(t) - r_0(t)| \rangle \qquad (6)$$

where $D$ is the diffusion coefficient, $r_i(t)$ is the coordinate of the center of the mass of the *i*th $H_2O$ molecule and $N$ is the number of calculated molecules in the system. The value of mean square displacement (MSD) of molecules calculated in MS is the average over a time interval for all molecules in a set. Therefore, **Equation 6** can be simplified to $D = a/6$, where $a$ denotes the slope of the best-fit line of MSD versus time.

The interactions of molecules and ions in IER were examined by calculating the radial distribution functions (RDFs) of atoms of interest. These functions, also referred to as pair correlation functions, provide insights into the structure of studied models. In a cell with volume V, for two groups of atoms A and B, they can be determined by **Equation 7**

$$g_{AB}(r) = \frac{V \times \langle \sum_{i \neq j} \delta(r - |r_{Ai} - r_{Bj}|) \rangle}{(N_A N_B - N_{AB}) 4\pi r^2 dr} \tag{7}$$

Where $i$ and $j$ refer to the $i$th and $j$th atoms in group A and group B. $N_{AB}$ is the number of atoms in both groups A and B, and the angle bracket implies averaging over different configurations. For a single group of atoms, accordingly, **Equation 7** can be simplified as

$$g_{AB}(r) = \frac{V \times \langle \sum_{i \neq j} \delta(r - |r_{ij}|) \rangle}{(N^2 - N) 4\pi r^2 dr} \tag{8}$$

these functions give the probability of finding an atom at a distance $r$ from another in a completely random distribution. They may be employed to investigate the interactions between quaternary ammonium cations and ions under different humidity conditions, and therefore, to analyze the role of water in IER systems.

## 3 Results and Discussion

### 3.1 Humidity Dependence of Diffusivity

Molecules diffusions in IER were studied under three relative humidity conditions (40%, 50%, 60%). The one-to-one corresponding humidity condintons to the ratios of $CO_3^{2-}$ : $H_2O$ are 1:5, 1:10 and 1:15 (S1), and the ratios of $HCO_3^-$ : $H_2O$ are 1:4, 1:9 and 1:14 (S2), respectively[1]. **Figure 5** shows the time-averaged MSDs of water molecules in S1 and S2 versus time with respect to the different humidity conditions. **Figure 6** shows the time-averaged MSDs of $CO_3^{2-}$ ions in S1, and $HCO_3^-$, $OH^-$ ions in S2 versus time with respect to different humidity conditions. Here, the slope of the plots are proportional to the diffusion coefficient in the system, therefore, the diffusion coefficients (D) were calculated. The diffusion coefficients of water molecules and ion species are shown in **Figure 7**.

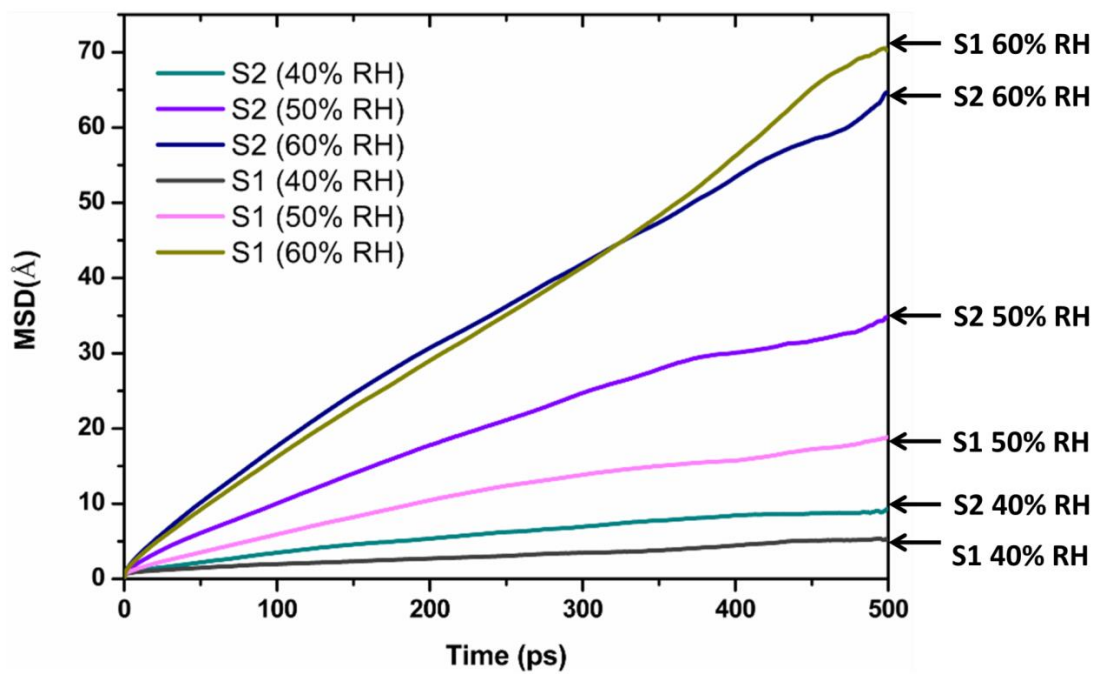

**Figure 5:** MSDs of water molecules in S1 and S2 versus time with respect to different humidity conditions.

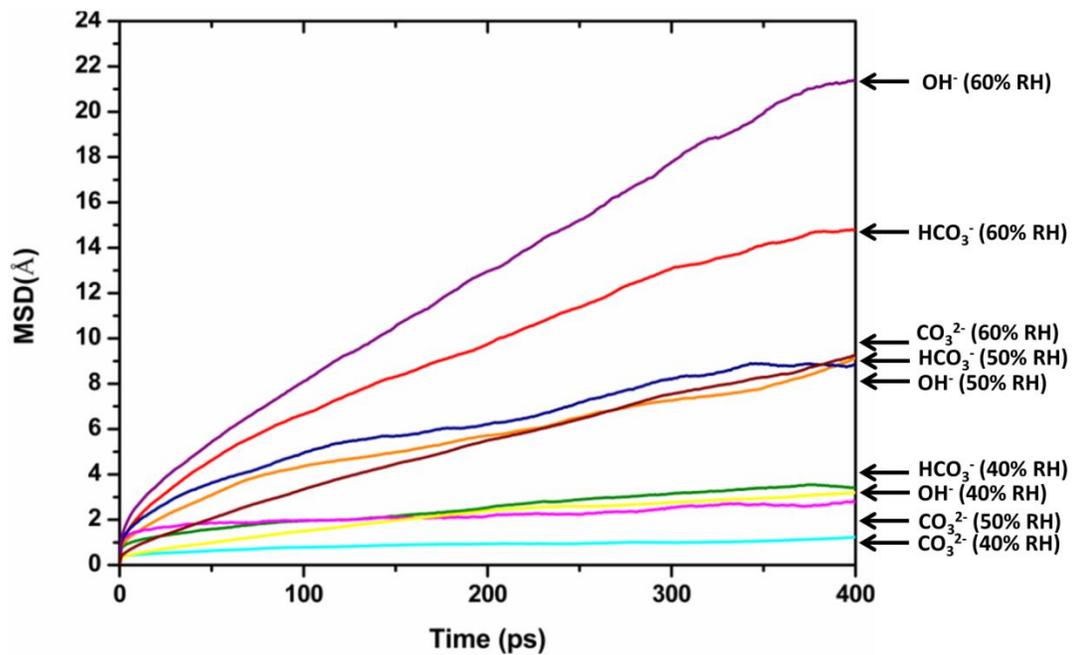

**Figure 6:** MSDs of ion $CO_3^{2-}$ in S1, and ions $HCO_3^-$, $OH^-$ in S2 versus time with respect to different humidity conditions.

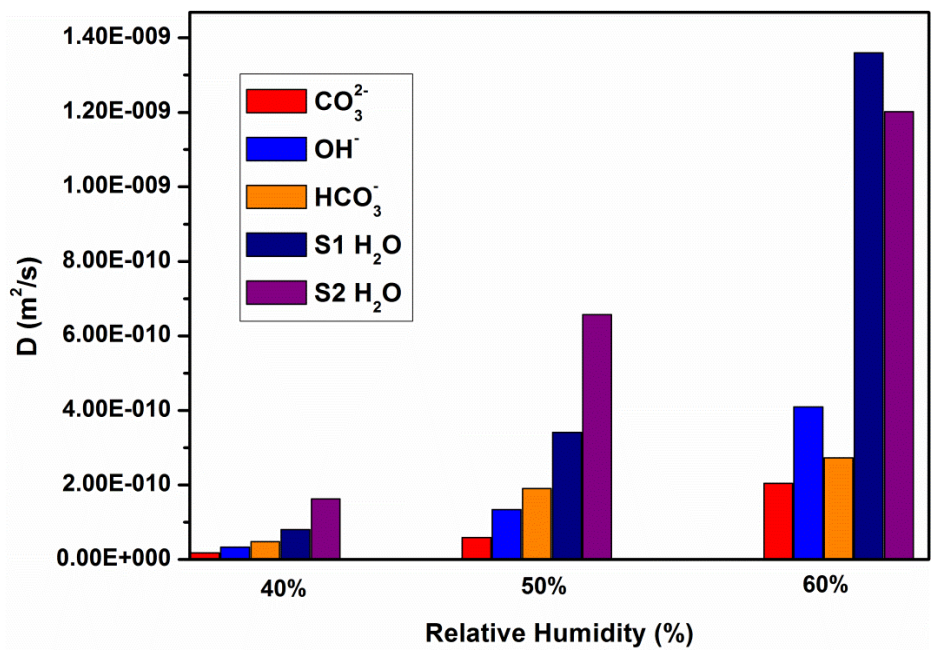

**Figure 7:** Diffusion coefficients of water molecules and ion species at various humidity conditions.

In general, the diffusion coefficients of water molecules and all ion species increase as the humidity increases. The higher humidity level means a higher level of hydration, which involves more water molecules which are uncoordinated to $NR_4^+$ groups, and these waters can break $NR_4^+$-$CO_3^{2-}$, $NR_4^+$-$HCO_3^-$, and $NR_4^+$-$OH^-$ pairs. Higher water content also leads to a better connected water-channel network, which can also stimulate water transport.

From a comparison of different ion species and water molecules under the same humidity condition, it is clear that the motilities of ion species are much lower than those of water molecules. This is reasonable, considering the strong electrostatic attraction between negative charged ions and $NR_4^+$ end-groups. Note that although $NR_4^+$ groups can exhibit local mobility, they are attached to the polystyrene backbone and therefore do not diffuse through the system. By comparison, water is much more unrestricted to move around in the simulation box, although water molecules can create relative weak H-bonds to $NR_4^+$ groups. Based on observation of local dynamics in the simulations, the water molecules and ion species which are farther away from backbones with $NR_4^+$ groups are much more mobile than those situated closer to the walls of the backbones. The quaternary ammonium cations $NR_4^+$ tend to immobilize and stabilize water molecules and ion species and thus reduce local mobility.

From the comparison of the diffusion coefficient of ion species in two systems, the results show the mobility of carbonate ions is lower than those of bicarbonate ions and also hydroxide ions, for all humidity levels considered herein. The reason is that the carbonate ion has higher valence which leads to a larger columbic force with $NR_4^+$ groups. The diffusion coefficient of bicarbonate ions is slightly higher than that of hydroxide ions under the 40% and 50% relative humidity conditions, shown in **Figure 7**. The reason is that the bicarbonate ions have a larger van der Waals force with backbone systems than hydroxide ions, which tends to immobilize bicarbonate ions more than hydroxide ions. However, when the relative humidity level raises to 60%, the mobility of hydroxide ion is higher than that of bicarbonate ions. Since the size of hydroxide ion is smaller than bicarbonate ion, the well-formed hydration shell may be created first when the water amount is up to a certain level, and then these floating ion hydrations could increase the mobility of hydroxide ions significantly.

## 3.2 Structure of Molecular System

The intermolecular RDFs for two pairs of atoms under different humidity conditions ($CO_3^{2-}$:$H_2O$ = 1:70, $CO_3^{2-}$:$H_2O$ = 1:50, $CO_3^{2-}$:$H_2O$ = 1:30, $CO_3^{2-}$:$H_2O$ = 1:10) are shown in **Figure 8**. One is nitrogen atoms in $NR_4^+$ and carbon atoms in $CO_3^{2-}$ (N-C), the other one is nitrogen atoms in $NR_4^+$ and carbon atoms in $HCO_3^-$ (N-C). The difference of this model from the one of diffusion analysis, is all anions ($CO_3^{2-}$, $HCO_3^-$, $OH^-$) are built in a single system with different water molecules. The ratio of $CO_3^{2-}$:$HCO_3^-$:$OH^-$ are 1:1:1 under the condition of neutral balance. There are indications from the molecular modeling, that the change of the number of water molecules can influence the precipitation rates of ionic species on the solid surfaces of the IER.

In a wet surrounding ($CO_3^{2-}$:$H_2O$ = 1:70), the hydration clouds of all ions are so large such that the anions can hardly approach the cations $RH_4^+$, shown in **Figure 8(a)** and **Figure 8(b)**. The probabilities of the appearance of $CO_3^{2-}$ and $HCO_3^-$ ions near the $NR_4^+$ groups are a little higher than that averaged in the whole system, because the atomic forces between $NR_4^+$ groups and anions are larger than the ones between water molecules. The probability of appearance of $HCO_3^-$ ions in the vicinity of $RH_4^+$ cations are higher than $CO_3^{2-}$ ions.

With less water molecules ($CO_3^{2-}$:$H_2O$ = 1:30), the optimal approach distances of $CO_3^{2-}$-$NR_4^+$ and $HCO_3^-$-$NR_4^+$ both become smaller. This, in turn, gives an energetic advantage to the mono-valent $HCO_3^-$ ion over the divalent $CO_3^{2-}$ ion that does not match the $NR_4^+$ single cationic charge. The presence of competition for one $CO_3^{2-}$ ion is between the two monovalent $NR_4^+$ cations. Therefore, the $NR_4^+$ quaternary ammonium cations do not comfortably accommodate carbonate ions into the structure. The part of unfitted $CO_3^{2-}$ ions containing hydration water are more likely to be located at 4.3 Å and 7.0 Å. On the other hand bicarbonate ions can easily fit to $NR_4^+$ cations and the favorable distance between them is at 4.5 Å, shown as **Figure 8(c)**. As a result, as the humidity level decreases, the $HCO_3^-$ ions are more likely to precipitate than $CO_3^{2-}$ ions, and favors a more alkaline surrounding. The larger amount of $OH^-$ ions in dry condition is more conducive to capture $CO_2$. This discovery may provide an explanation for the underlying mechanism of IER absorbs $CO_2$ in dry and release $CO_2$ in wet.

At even lower humidity ($CO_3^{2-}:H_2O = 1:10$), both carbonate ions and bicarbonate ions precipitate completely. One obvious peak shows at distance of 4.5 Å, shown as **Figure 8(d)**. The sharp peak at distance around 4.5 Å is an indication of the strong columbic force and van der Walls force of $NR_4^+$ cations associated with $CO_3^{2-}$ anions in S1 or $HCO_3^-$ anions in S2. The attracted $CO_3^{2-}$ anions and $HCO_3^-$ anions are more likely to be located in the vicinity of the $NR_4^+$ groups on the IER network in the dry condition. Under this condition, the mechanism of $CO_2$ capture by IER was explained elsewhere[1]. The intermolecular RDFs for two pairs of atoms under more humidity conditions ($CO_3^{2-}:H_2O = 1:5$ to $CO_3^{2-}:H_2O = 1:70$) are provided in supplementary materials.

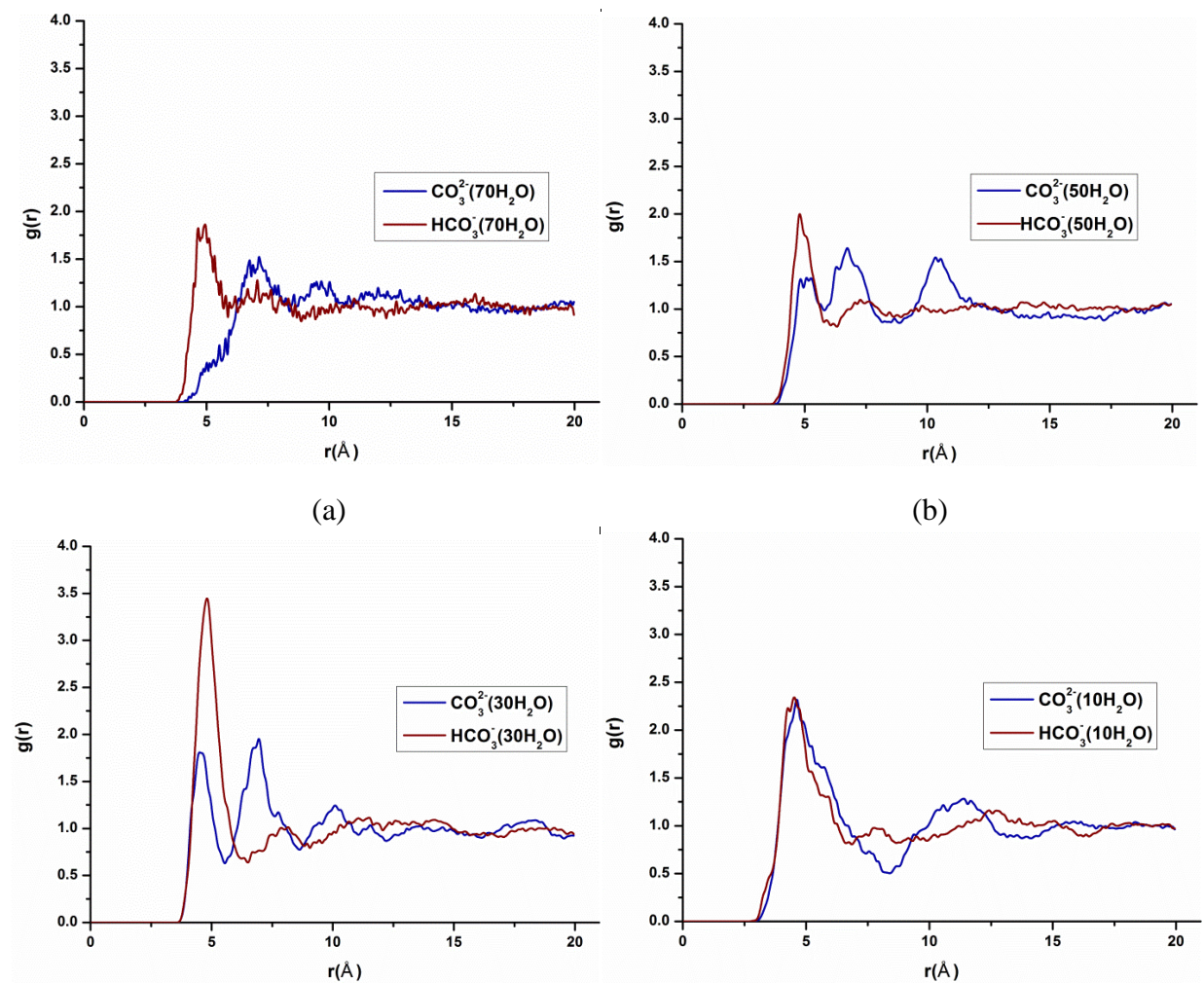

(a)  (b)

(c)  (d)

(c)                                                    (d)

**Figure 8:** Intermolecular radial distribution functions between 1) Navy color: N atoms in $NR_4^+$ and C atoms in $CO_3^{2-}$. 2) Wine color: N atoms in $NR_4^+$ and C atoms in $HCO_3^-$. These two RDFs are calculated under three humidity conditions: a) $CO_3^{2-}:H_2O = 1:70$, b) $CO_3^{2-}:H_2O = 1:50$, c) $CO_3^{2-}:H_2O = 1:30$, and d) $CO_3^{2-}:H_2O = 1:10$

# 4 Summary

The work reports the results of MD simulations of moisture-swing $CO_2$ sorbent with carbonate ion system and bicarbonate ion system under different humidity conditions. The transport characteristics and structures of ions species are explored with different numbers of water molecules. The diffusion coefficients of water molecules and anions provide helpful insights, from a molecular level perspective, for designing a $CO_2$ sorbent with better dynamic performance. The $CO_2$ capture efficiency can be enhanced according to increase the ion diffusion rates, which could be realized by using different support materials with different characteristics like hydrophobicity and cation species. The molecular structure analysis states the different precipitation rates of carbonate ions and bicarbonate ions. In a drier surrounding, bicarbonate ion that precipitates out first leaves behind a more alkaline solution, which may promote the absorption of $CO_2$. This finding may provide an elucidation for the working mechanism of moisture-swing $CO_2$ sorbent.

1   Shi, X., Xiao, H., Lackner, K. S. & Chen, X. Capture CO2 from Ambient Air Using Nanoconfined Ion Hydration. *Angew. Chem.* (2016).
2   Mishra, H. *et al.* Brønsted basicity of the air–water interface. *Proceedings of the National Academy of Sciences* **109**, 18679-18683, doi:10.1073/pnas.1209307109 (2012).

3      Kunz, W., Henle, J. & Ninham, B. W. 'Zur Lehre von der Wirkung der Salze' (about the science of the effect of salts): Franz Hofmeister's historical papers. *Current Opinion in Colloid & Interface Science* **9**, 19-37, doi:10.1016/j.cocis.2004.05.005 (2004).

4      Jungwirth, P. & Tobias, D. J. Specific Ion Effects at the Air/Water Interface. *Chem. Rev.* **106**, 1259-1281 (2005).

5      Zidi, Z. S. Solvation of Sodium-chloride Ion Pair in Water Cluster at Atmospheric Conditions: Grand Canonical Ensemble Monte Carlo Simulation. *The Journal of Chemical Physics* **123**, doi:10.1063/1.1979476 (2005).

6      Zhao, Z., Rogers, D. M. & Beck, T. L. Polarization and Charge Transfer in the Hydration of Chloride Ions. *The Journal of Chemical Physics* **132**, doi:10.1063/1.3283900 (2010).

7      Shapiro, N. & Vigalok, A. Highly Efficient Organic Reactions "on Water", "in Water", and Both. *Angew. Chem. Int. Ed.* **47**, 2849-2852, doi:10.1002/anie.200705347 (2008).

8      Poynor, A. *et al.* How Water Meets a Hydrophobic Surface. *Phys. Rev. Lett.* **97**, 266101 (2006).

9      Xiao, H. *et al.* The catalytic effect of water in basic hydrolysis of $CO_3{}^{2-}$ in hydrated clusters. *arXiv preprint arXiv:1702.03036* (2017).

10     Li, Q. *et al.* Molecular characteristics of H2O in hydrate/ice/liquid water mixture. *Int. J. Mod Phys B* **29**, 1550185, doi:10.1142/s0217979215501854 (2015).

11     Lackner, K. S. Capture of Carbon Dioxide from Ambient Air. *Eur. Phys. J. Spec. Top.* **176**, 93-106, doi:10.1140/epjst/e2009-01150-3 (2009).

12     Doherty, D., Holmes, B., Leung, P. & Ross, R. Polymerization molecular dynamics simulations. I. Cross-linked atomistic models for poly (methacrylate) networks. *Comput. Theor. Polym. Sci.* **8**, 169-178 (1998).

13     Yarovsky, I. & Evans, E. Computer simulation of structure and properties of crosslinked polymers: application to epoxy resins. *Polymer* **43**, 963-969 (2002).

14     Heine, D. R., Grest, G. S., Lorenz, C. D., Tsige, M. & Stevens, M. J. Atomistic simulations of end-linked poly (dimethylsiloxane) networks: structure and relaxation. *Macromolecules* **37**, 3857-3864 (2004).

15     Varshney, V., Patnaik, S. S., Roy, A. K. & Farmer, B. L. A molecular dynamics study of epoxy-based networks: cross-linking procedure and prediction of molecular and material properties. *Macromolecules* **41**, 6837-6842 (2008).

16     Liu, J. *et al.* Understanding flocculation mechanism of graphene oxide for organic dyes from water: Experimental and molecular dynamics simulation. *AIP Advances* **5**, 117151, doi:doi:http://dx.doi.org/10.1063/1.4936846 (2015).

17     Stillinger, F. H. & Rahman, A. Molecular dynamics study of temperature effects on water structure and kinetics. *The Journal of Chemical Physics* **57**, 1281-1292 (1972).

18     Matsumoto, M., Saito, S. & Ohmine, I. Molecular dynamics simulation of the ice nucleation and growth process leading to water freezing. *Nature* **416**, 409-413 (2002).

19     Bharadwaj, R. K. & Boyd, R. H. Small molecule penetrant diffusion in aromatic polyesters: a molecular dynamics simulation study. *Polymer* **40**, 4229-4236 (1999).

20     Liu, J. W., Mackay, M. E. & Duxbury, P. M. Molecular Dynamics Simulation of Intramolecular Cross-Linking of BCB/Styrene Copolymers. *Macromolecules* **42**, 8534-8542, doi:10.1021/ma901486q (2009).

21     Lu, C., Ni, S., Chen, W., Liao, J. & Zhang, C. A molecular modeling study on small molecule gas transportation in poly (chloro-p-xylylene). *Computational Materials Science* **49**, S65-S69 (2010).

| 22 | Tsuzuki, S., Uchimaru, T., Mikami, M. & Urata, S. Magnitude and orientation dependence of intermolecular interaction of perfluoropropane dimer studied by high-level ab initio calculations: Comparison with propane dimer. *The Journal of chemical physics* **121**, 9917-9924 (2004). |
|---|---|
| 23 | Pavel, D. & Shanks, R. Molecular dynamics simulation of diffusion of O 2 and CO 2 in blends of amorphous poly (ethylene terephthalate) and related polyesters. *Polymer* **46**, 6135-6147 (2005). |
| 24 | Lin, Y. & Chen, X. Investigation of moisture diffusion in epoxy system: experiments and molecular dynamics simulations. *Chem. Phys. Lett.* **412**, 322-326 (2005). |
| 25 | Wu, C. & Xu, W. Atomistic simulation study of absorbed water influence on structure and properties of crosslinked epoxy resin. *Polymer* **48**, 5440-5448 (2007). |
| 26 | Chang, S.-H. & Kim, H.-S. Investigation of hygroscopic properties in electronic packages using molecular dynamics simulation. *Polymer* **52**, 3437-3442 (2011). |
| 27 | Lee, S. G., Jang, S. S., Kim, J. & Kim, G. Distribution and diffusion of water in model epoxy molding compound: molecular dynamics simulation approach. *Advanced Packaging, IEEE Transactions on* **33**, 333-339 (2010). |
| 28 | Humbert, H., Gallard, H., Suty, H. & Croué, J.-P. Performance of selected anion exchange resins for the treatment of a high DOC content surface water. *Water Res.* **39**, 1699-1708 (2005). |
| 29 | Johnson, C. J. & Singer, P. C. Impact of a magnetic ion exchange resin on ozone demand and bromate formation during drinking water treatment. *Water Res.* **38**, 3738-3750 (2004). |
| 30 | Feng, D., Aldrich, C. & Tan, H. Treatment of acid mine water by use of heavy metal precipitation and ion exchange. *Miner. Eng.* **13**, 623-642 (2000). |
| 31 | Wang, T., Lackner, K. S. & Wright, A. B. Moisture-Swing Sorption for Carbon Dioxide Capture from Ambient Air: A Thermodynamic Analysis. *PCCP* **15**, 504-514, doi:10.1039/c2cp43124f (2013). |
| 32 | Wang, T. *et al.* Characterization of kinetic limitations to atmospheric CO2 capture by solid sorbent. *Greenhouse Gases: Science and Technology* (2015). |
| 33 | Shi, X., Xiao, H., Chen, X. & Lackner, K. S. The Effect of Moisture on the Hydrolysis of Basic Salts. *Chemistry-A European Journal* **22**, 18326-18330 (2016). |
| 34 | Shi, X., Xiao, H., Chen, X. & Lackner, K. A Carbon Dioxide Absorption System Driven by Water Quantity. *arXiv preprint arXiv:1702.00388* (2017). |
| 35 | Brandell, D., Karo, J., Liivat, A. & Thomas, J. O. Molecular dynamics studies of the Nafion®, Dow® and Aciplex® fuel-cell polymer membrane systems. *J. Mol. Model.* **13**, 1039-1046 (2007). |
| 36 | http://accelrys.com/products/materials-studio/. |
| 37 | Einstein, A. *Investigations on the Theory of the Brownian Movement*. (Courier Corporation, 1956). |